\documentclass[11pt]{article}
\usepackage[utf8]{inputenc}
\usepackage{hyperref}
\usepackage[margin=1in]{geometry}
\newgeometry{vmargin={2cm}, hmargin={20mm,20mm}}
\usepackage[width=0.9\textwidth,font=footnotesize,labelfont=bf,]{caption}
\usepackage{amsmath}
\usepackage{amssymb}
\usepackage{todonotes}
\usepackage{subcaption}
\usepackage[capitalise]{cleveref}
\RequirePackage[numbers, sort&compress]{natbib}
\usepackage{siunitx}

\newcommand{\cc}{0.5\% MC-PBS}
\newcommand{\ccb}{0.6\% MC-PBS}

\title{A New Hyperelastic Lookup Table for RT-DC}

\author{Lucas D. Wittwer$^{1,2,3}$, Felix Reichel$^{3}$, Paul M{\"u}ller$^{3}$, Jochen Guck$^{3}$, Sebastian Aland$^{1,2}$}

\date{
$^1$Numerical Mathematics and Optimisation, Freiberg University of Mining and Technology, Akademiestrasse 6, 09599 Freiberg, Germany\\
$^2$Faculty of Informatics/Mathematics, University of Applied Sciences, Friedrich-List-Platz 1, 01069 Dresden, Germany\\
$^3$MPL \& MPZPM Erlangen, Staudtstrasse 2, 91058 Erlangen, Germany\\
\today
}

\begin{document}

\maketitle
\begin{abstract}
Real-time deformability cytometry (RT-DC) is an established method that quantifies features like size, shape, and stiffness for whole cell populations on a single-cell level in real time.
To extract the cell stiffness, a lookup table (LUT) disentangles the experimentally derived steady state cell deformation and the projected area, yielding the Young's modulus.
So far, two lookup tables exists, but are limited to simple linear material models and cylindrical channel geometries.
Here, we present two new lookup tables for RT-DC based on a neo-Hookean hyperelastic material numerically derived by simulations based on the finite element method in square and cylindrical channel geometries.
At the same time, we quantify the influence of the shear-thinning behaviour of the surrounding medium on the stationary deformation of cells in RT-DC and discuss the applicability and impact of the proposed LUTs regarding past and future RT-DC data analysis.
Additionally, we provide insights about the cell strain and stresses, as well as the influence resulting from the rotational symmetric assumption on the cell deformation and volume estimation.
The new lookup tables as well as the numerical cell shapes are made freely available.
\end{abstract}


\section*{Background}

Cell stiffness emerged as an important phenotypical marker that is linked to a multitude of biological processes and can reveal pathological changes of cells \citep{Toepfner2018-al, Koch2017-ji}.
The emergence of high-throughput techniques to measure single cell deformability enables the use of cell deformation as a tool in clinical diagnostics \citep{Kubankova2021-uu}.
In this work, we concentrate on the microfluidic technique \emph{real-time deformability cytometry} (RT-DC) \citep{Otto2015-hf}.
In RT-DC, cells flow through a narrow microfluidic channel where they get deformed by hydrodynamic stresses from the flow field around them, reaching a steady state configuration.
The deformed cells are imaged, and the contour is analysed in real time at a throughput of up to 1000 cells/s.
Based on this stationary deformation, not only base properties like the cell circumference $L$, projected area~$A$ and volume $V$ are deduced but also material properties like the cell's stiffness (elasticity) in the form of the Young's modulus.
For the latter, the influence of the cell size, measured by the area $A$, needs to be decoupled from the stationary deformation $D$, here defined as 
\begin{align}
	D = 1 - \text{circularity} = 1 - \frac{2 \sqrt{\pi A}}{L},
\end{align}
which is a measure for the deviation of a shape from a circle ($ \text{circularity} = 1$).
Analytical \citep{Mietke2015-cj} and numerical models \citep{Mokbel2017-pb} were developed to generate \emph{lookup tables} (LUT) mapping an area-deformation pair to a unique apparent Young's modulus~$E$.

RT-DC measurements are done for various flow rates, channel widths and measurement buffer viscosities. 
Mietke et al.\ 2015 showed analytically that isoelasticity lines -- the level set for a fixed Young's modulus -- of the lookup table can be scaled to a different channel width $L^\prime$, flow rate $Q^\prime$ and viscosity of the buffer medium $\eta^\prime$ for a Newtonian carrier and linear elastic cell material by $E' = E \left(\eta^\prime Q^\prime L^3 / \eta Q L^{\prime 3}\right)$ and $A^\prime = A \left(L^\prime / L \right)^2$.
Here, $\eta$, $Q$ and $L$ are the parameters used to generate the lookup tables.
Thus, the same lookup table can be used for different flow rates, microfluidic channels and buffers. 
Those lookup tables as well as the scaling are implemented in \emph{dclab} \cite{dclab}, the Python library for post-measurement analysis of real-time deformability cytometry.

However, these two models use three simplifications. 
First, the square cross-section of the channel is assumed to be cylindrical to employ rotational symmetry. Second, they do not consider the non-Newtonian shear-thinning nature of the suspension buffers used in RT-DC, e.g., methyl-cellulose dissolved in phosphate buffered saline (MC-PBS) \cite{Herold2017-ir}.
And last, both assume a small strain linear elasticity bulk model for the cell.
Especially for larger cells or higher hydrodynamic stresses, the small strain assumption does not hold.
Recent work on probing viscoelastic properties of cells in RT-DC, dynamic real-time deformability cytometry \emph{dRT-DC} \cite{Fregin2019-up}, showed large deformation in the inflow region of the narrow channel. 
A numerical study of the viscoelastic behaviour of cells and beads in RT-DC has been conducted in the axisymmetric case in \cite{Schuster2021-hx} and in full three-dimensional channels in \cite{Wittwer2022-mj}.
Both studies model the cell as a neo-Hookean hyperelastic material model.
The apparent viscosity $\eta$ is linked to the apparent Young's modulus by $\eta = E \tau$, where $\tau$ is the relaxation timescale of the cell. 
But $E$ is only deducible by a lookup table based on the same material model. 
Thus, a LUT based on the neo-Hookean hyperelastic material model is necessary to deduce the apparent Young's modulus $E$ and viscosity~$\eta$. 

Here, we introduce two new improved LUTs -- one for a cylindrical and one for a square channel -- to extract the stiffness of cells in RT-DC based on fully coupled fluid-solid interaction finite element simulations.
We use a hyperelastic neo-Hookean material to get the cells' elastic response in a full three-dimensional square channel with a shear-thinning buffer. 
We show that like the initial analytical and numerical LUTs, the two proposed LUTs uniquely map every combination of area and deformation to a Young's modulus and the concept of the scaling still holds, including shear-thinning. 
The new and improved LUT for RT-DC will help to reveal stiffness changes between cell types or due to cell state changes with yet unprecedented accuracy.
Additionally, we report for the first time the resulting strains and hydrodynamic surface stresses acting on the cells.
And finally, we investigate the systematic error in volume computation introduced by the widely-used assumption of rotationally symmetric cell deformation.
\section*{Methods}


We start by describing the computational domain and the new fluid-solid interaction (FSI) model with a shear-thinning non-Newtoinan fluid (the buffer) and a hyperelastic neo-Hookean solid (the cell). 
Based on this model, we describe the generation of the new lookup table by sampling the area-deformation space.
At the end of this section, we describe the experimental setup used to measure the HL60 cells and PAAm beads in \cref{fig:beads}.

\subsection*{Fluid-solid interaction model}
\begin{figure}[tp]
	\centering
	\includegraphics[width=.75\textwidth]{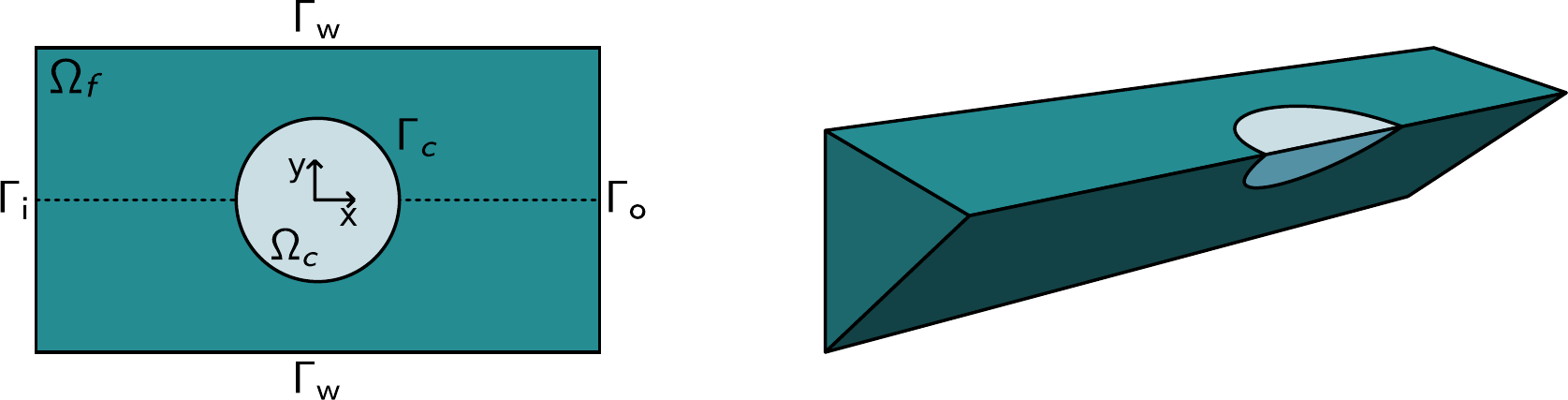}
	\caption{
		{\bf Computational domain:} 
		{\bf (left)}
		Schema of the computational domain (top view). 
		The origin of the coordinate system is in the centre of the cell, the z-axis is out of
		plane.
		The fluid domain $\Omega_f$ (darker shade) is enclosed by the inflow and outflow boundaries $\Gamma_i$ and $\Gamma_0$ respectively and the non-slip boundaries $\Gamma_w$.
		The fluid surrounds the cell domain $\Omega_c$ with the fluid-solid interface $\Gamma_c$.
		In the axisymmetric case, the upper half is considered only. 
		{\bf (right)}
		Schema of the full three-dimensional domain for the square channel simulations. 
		Using symmetries in the channel geometry, the computational domain can be reduced to $1/8$ of the actual domain, as indicated.
	}
	\label{fig:compdomain}
\end{figure}

The computational domain as illustrated in \cref{fig:compdomain} is split into the fluid subdomain $\Omega_f$ and the subdomain representing the cell $\Omega_c$. The subdomains are separated by the interface $\Gamma$. 
We model the fluid in the $\Omega_f$ by the incompressible Navier-Stokes equations given by
\begin{align}
    \rho_f \frac{\partial \mathbf{u}}{\partial t} + \rho_f \left(\mathbf{u} \cdot \nabla \right) \mathbf{u} &= \nabla \cdot \mathbf{\sigma}_f = \nabla \cdot \left[-p \mathbf{I} + \eta(\mathbf{u}) \left(\nabla \mathbf{u} + (\nabla \mathbf{u}^T)\right) \right] & \text{in} \ \Omega_\text{f}\\
    \rho_f \nabla \cdot \mathbf{u} &= 0 & \text{in} \ \Omega_\text{f}
\end{align}
with $\rho_f \in \mathbb{R}$ is the fluid density, $\mathbf{u} \in \mathbb{R}^3$ the fluid velocity, $p \in \mathbb{R}$ the pressure, and $\mathbf{I} \in \mathbb{R}^{3x3}$ the identity matrix.
The power-law rheology model with $\eta(\mathbf{u}) = m \left( \dot{\gamma} \right)^{n - 1} \in \mathbb{R}$ accounts for the velocity-dependent viscosity where $m \in \mathbb{R}$ is the fluid consistency coefficient, $\dot{\gamma}$ is the shear rate and $n \in \mathbb{R}$ the flow behaviour index.

Following \citep{Mokbel2017-pb}, we assume the cells to be incompressible but not linear elastic but as a non-linear neo-Hookean hyperelastic material described by
\begin{align}
    \rho_s \frac{\partial^2 \mathbf{w}}{\partial t^2} &= \nabla \cdot \left(\mathbf{\sigma}_c \right)^T = \nabla \cdot \left( \mathbf{F} S \right)^T &\text{in} \ \Omega_c
\end{align}
where $\rho_s \in \mathbb{R}$ is the cell density, $\mathbf{w} \in \mathbb{R}^3$ the displacement vector, 
$\mathbf{F}$ is the deformation gradient and $S$ is the second  Piola-Kirchhoff  stress. 
The strain energy density of the neo-Hookean material $W_s$ is
\begin{align}
    W_s &= \frac{1}{2} \mu \left( \bar{I}_1 - 3 \right) + \frac{1}{2} \kappa \left(J_{el} - 1 \right)^2
\end{align}
which is the sum of the isochoric strain  energy  density (using the isochoric invariant $\bar{I}_1$) and the volumetric strain energy density (using the elastic volumetric deformation $J_{el}$). 
In the case of an incompressible material, the parameters are given by the Lamé parameter $\mu = E / 3 \in \mathbb{R}$ with the Young's modulus $E$ and the bulk modulus $\kappa \in \mathbb{R}$.

On the interface $\Gamma$, we impose the kinematic condition $\mathbf{u} = \partial_t \mathbf{w}$ (continuity of velocities) as well as the dynamic condition $\mathbf{n}_f \cdot \mathbf{\sigma}_f = - \mathbf{n}_c \cdot \mathbf{\sigma}_c$ (balance of forces). 
The mechanical model is described in more details in \cite{Wittwer2022-mj}. 

\subsection*{Finite-Element Implementation}
The above system of equations is discretized and solved by the finite element method using COMSOL Multiphysics. 
For the full three-dimensional simulations of the square channel, we reduced the computational effort by exploiting the 4-fold rotational symmetry and 2-fold reflection symmetry of the rectangular channel cross-section (see \cref{fig:compdomain} on the right).
For the cylindrical channel lookup table, we can exploit the rotational symmetry and reduce the computational domain to a two-dimensional surface. 
The radius $R$ of the cylindrical channel is adapted to $R = 1.094 \times l_c$, where $l_c$ is the square channel width, such that the pressure drop over the cylindrical channel is equal to the square channel (see the concept of \emph{equivalent channel} described in \cite{Mietke2015-cj}).
The resulting computational domain is meshed with a combination of tetrahedral, prisms and pyramid elements to efficiently reduce the number of degrees of freedom even further. 
To resolve the high gradient in the flow profile on the channel walls and the cell surface, we added a boundary layer on $\Gamma_w$ and $\Gamma_c$. 
We choose linear Lagrange elements (P1) for discretizing the Navier-Stokes equation for both the velocity field and pressure and stabilize the saddle-point problem with streamline and crosswind diffusion.
Linear elements are used to discretize the displacement field of the hyperelastic material. 
We did not experience any locking behaviour during our model validation by comparing to the solution with quadratic elements for the solid. 

The correct inflow profile $\mathbf{u}_i$ is precomputed in the absence of a cell, based on the flow rate, channel width and the non-Newtonian parameters of the fluid and used as Dirichlet boundary condition at the inlet. 
At the outlet $\Gamma_o$, the pressure is fixed to $p_o = 0$ and we suppressed any eventual backflow. 
To keep the cell centred in the computational domain, we describe all quantities in a coordinate system which moves with the cell. 
We use a proportional integral derivative (PID) controller to adjust the wall speed $\mathbf{u}_w$ on $\Gamma_w$ based on the barycentre of the cell. 
This wall speed is subtracted from the inflow profile such that the resulting system is the same as if the cell moves through the channel. 
The arbitrary Lagrangian-Eulerian (ALE) description is employed for the deformation of the fluid domain $\Omega_f$. 
The grid movement on $\Gamma$ is extended into the interior of the domain and smoothed by treating the vertex displacements as a neo-Hookean hyperelastic material. 

To improve the stability of the simulations, we add an artificial viscosity to the cell in a Kelvin-Voigt like manner, and we monotonically ramp up the inflow profile from $\mathbf{u} = 0$ to $\mathbf{u}_\text{i}$. We stop the simulation as soon as the steady state deformation is reached by measuring the absolute deformation change $\frac{\partial d}{\partial t}$ and set as the stop criteria $\frac{\partial d}{\partial t} \leq \num{1e-3}$. The steady state is not influenced by the viscous term and the ramp up.
The fully bidirectional coupled geometrical non-linear system is solved with a Newton method in time and the PARDISO direct solver in space.

\subsection*{Creation of the new lookup tables}
We sampled the area-deformation space for the new lookup tables indirectly by the radius and  Young's modulus of the cells. 
The radii $R$ are chosen equidistant in $R^2$ such that the resulting projected areas should be in the range of $[\SI{30}{\um^2}, \SI{290}{\um^2}]$ (the actual projected cell area is not conserved). 
Similar, the Young's modulus $E \in [\SI{0.3}{kPa}, \SI{30}{kPa}]$ is equally sampled in $E^{-1/2}$.
For each $R \times E$ combination, we ran two simulations, one in a square channel with a channel width of $l_c = \SI{20}{\um}$ and one in an equivalent cylindrical channel (see above). 
The cell has a density of $\rho_s = \SI{1000}{kg.m^{-3}}$ and we set the bulk modulus $\kappa = \SI{2.15}{GPa}$ corresponding to the bulk modulus of water, leading to a quasi-incompressible cell.
The artificial cell viscosity is set to $\SI{10}{mPa.s}$. 
For both lookup tables, the flow rate is $\SI{0.04}{\ul.s^{-1}}$.
The fluid parameters are based on \ccb{} (see below). 
Additionally, we set the density to $\rho_f = \SI{1000}{kg.m^{-3}}$.

For both lookup tables, we did several steps of data cleaning to make sure that the simulations did converge and are not numerical artefacts.
Based on polynomial regression and the random sample consensus (RANSAC) method, we automatically detect and delete outliers. 
Additionally, we drop all the simulations with a deformation $D \leq 5e-4$, since such deformation values are not measurable by RT-DC and are close to the perfect sphere discretized by the mesh. 
The final lookup tables are built by linearly interpolating the 22'558 obtained cell shapes for the cylindrical (2D axisymmetric) channel and 1'206 cell shapes for the square (3D) channel to a uniform grid for fast lookup. 

\subsection*{Cell culture}
The HL60/S4 cell subline (ATCC Cat\# CRL-3306, RRID:CVCL\_II77) was cultured in RPMI 1640 medium with \SI{2}{mM} L-Glutamine (Thermo Fisher \#A1049101) with 1\% penicillin and streptomycin (Gibco) and 10\% heat-inactivated fetal bovine serum (Sigma Aldrich, catalogue no. F4135, lot no. 13C519). Cells were grown at \SI{37}{\celsius}, with 5\% CO2, at densities between $10^5 - 10^6$ cells per ml with subculturing every $48 - 72$ hours.

\subsection*{Buffer production for RT-DC}
Samples for RT-DC experiments were suspended in a buffer solution made from 0.594\,w/w\% methyl cellulose (MC; 4,000 cPs, Alfa Aesar 036,718.22, CAS\#9,004–67–5) dissolved in phosphate buffered saline (PBS) without Mg$^{2+}$ and Ca$^{2+}$ (Gibco Dulbecco 14190144). Buffers were adjusted to have a viscosity of \SI{25}{mPa.s} when measured in a HAAKE Falling Ball Viscometer type C (Thermo Fisher Scientific, Dreieich, Germany) using ball number 3 at \SI{24}{\celsius}.

The rheology of the 0.6\% MC-PBS solutions was measured with an Anton Paar MCR 502WESP TwinDrive rheometer (Anton Paar GmbH, Graz, Austria) at \SI{25}{\celsius} controlled with a PTD 180 MD Peltier element (Anton Paar GmbH, Graz, Austria). At shear rates over \SI{5.000}{s^{-1}}, the solutions showed power law behaviour and the fluid consistency coefficient was found at $m=\SI{0.4057}{Pa.s}$ and the flow behaviour index at $n=0.6039$.
For the \cc{} buffer in the scaling validation, we used the measured fluid consistency coefficient of $\SI{0.2671}{Pa.s}$ and a flow behaviour index of $0.6264$.

\subsection*{Cell and Beads Measurements}
Cell and beads experiments were performed with a commercially available RT-DC device (Accelerator, Zellmechanik Dresden GmbH). The measurement principle of RT-DC is described in detail in Otto et al. \citep{Otto2015-hf}. Briefly, a microfluidic chip containing the measurement channels, that were simulated in this work, is mounted on an inverted microscope (Axiovert 200M, ZEISS, Oberkochen, Germany) and flow is introduced with syringe pumps. Images are captured with a CMOS-camera (Mikrotron, Unterschleissheim, Germany). Syringe pump and camera were controlled with the measurement software ShapeIn (Zellmechanik Dresden, Dresden, Germany) which analyses contours in real time.

For the beads experiments, \SI{2.5}{\ul} of beads solution (for details, see Girardo et al. \citep{Girardo2018-yk}) were resuspended in \SI{47.5}{\ul} of 0.6\% MC-PBS and measured at a flow rate of \SI{0.04}{\ul.s^{-1}}. 

For cell experiments, \SI{2}{ml} of HL60-cell solution was centrifuged at \SI{188}{RCF} for \SI{4}{minutes}. After the supernatant was removed, the cell pellet was resuspended in \SI{100}{\micro l} of 0.6\% MC-PBS and measured at \SI{0.04}{\micro l.s^{-1}}.

In both cases, the measurements have been pixelation corrected (see \cite{Herold2017-ir} for more details) and only cells and beads with an area-ratio of $1.0 - 1.05$ are considered \cite{Urbanska2018-hs, Herbig2018-gl}. 

\section*{Results}
The new lookup tables are based on finite element simulations, where we indirectly sample the area-deformation space by varying the cell radius and Young's modulus. 
Thereby, we get for each fixed Young's modulus an \emph{isoelasticity line} (see \cite{Mietke2015-cj}) in the area-deformation space. 
We commence by describing one of these numerically obtained cell shapes and show the flow field around \emph{in-silico} cell, as well as the surface stresses acting on it.
Next, we validate the new LUTs by comparing to the linear elastic lookup table from \cite{Mokbel2017-pb} and show that the scaling still holds. 
We compare the new lookup tables in the cylindrical and square channel to the previous lookup table.
Following this, we show the influence of the different lookup tables on the resulting apparent Young's modulus.
We conclude this section by reporting the engineering strains and surfaces stresses on biological cells and the relative volume error resulting from the axisymmetric channel assumption in the steady state configuration.
 
\subsection*{Cell Shape and Stress Distribution}
\begin{figure}
    \centering
    \includegraphics[width=.75\textwidth]{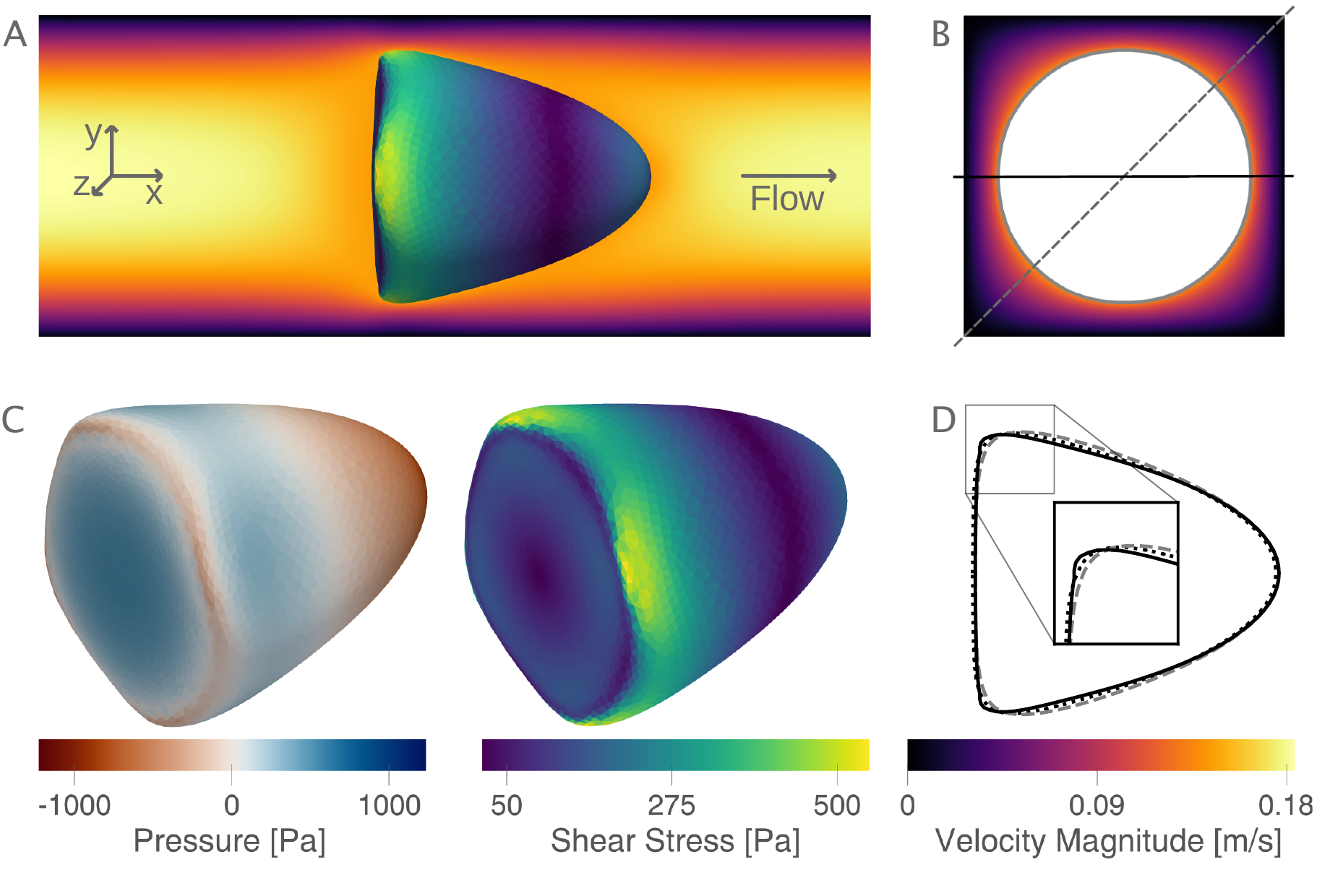} 
	\caption{
	    {\bf Cell shape, flow field, stress distribution and cell contours:} 
		{\bf A)} 
		Magnitude of the flow field (left to right) in a centred cross-section of the channel ($z = \SI{0}{\um}$) with a deformed cell (top view).
		{\bf B)} 
		shows the flow magnitude perpendicular to the flow direction.
		{\bf C)} 
		The pressure and shear stress on the cell surface.
		The pressure is normalized by subtracting the average surface pressure.
		{\bf D)} 
		The corresponding contour of the cell (solid line). 
		The dashed line is the contour in the diagonal, as indicated in {\bf B)}. 
		The dotted line is the same cell deformed in the equivalent cylindrical channel.
		Parameters: 
		$r = \SI{7.82}{\um}$,
		$E = \SI{1}{kPa}$,
		$L = \SI{20}{\um}$,
		$Q = \SI{0.04}{\ul.s^{-1}}$,
		buffer characteristics of \ccb{}
		}
	\label{fig:overview}
\end{figure}

The new proposed LUTs assume a neo-Hookean hyperelastic material for biological cells and are deduced from full three-dimensional finite element simulations and two-dimensional axisymmetric simulations.  
In \cref{fig:overview}, we show one of the three-dimensional simulations.
The \emph{in-silico} cell reaches a bullet-shaped stationary deformation (see \cref{fig:overview}A) due to the hydrodynamic stresses from the surrounding fluid.
The stationary configuration of the cell is not rotationally symmetric.
The flow profile around the cell from the front is shown in \cref{fig:overview}B.
The stresses on the cell surface can be split in normal (pressure) and tangential (shear) stress contribution.  
\Cref{fig:overview}C shows the pressure on the cell surfaces and the shear stress distribution.
We subtract the average surface pressure $p_\text{avg} = \int_A p \ dA / \int_A 1 \ dA$ since the pressure enters the system as a gradient only and, thus, the absolute value is not meaningful.
The pressure is non-symmetric, with the highest absolute value at the front of the cell. 
Here, we define the shear stress $||\mathbf{P} \left(\mathbf{\sigma}_c \cdot \mathbf{n} \right)||$ where $\mathbf{P} = \mathbf{I} - \mathbf{n} \otimes \mathbf{n}$ is the surface projection, $\sigma_c$ the stress tensor of the cell and $\mathbf{n}$ the surface normal).
The peak shear stress is located in the regions closest to the channel wall.

Only the projected area and deformation of the resulting cell shape are used for the lookup creation. 
In \cref{fig:overview}D, we plot this projected contour for the cylindrical channel overlaid with the projected contour in the square channel.
The corresponding deformation values are $D_\text{cylindrical} = 0.071$ and $D_\text{square} = 0.086$.
For comparison, the contour of the cell along the diagonal is shown with a dashed line. 

\subsection*{Model and Scaling Validation}
\begin{figure}[tp]
    \centering
    \includegraphics[width=.75\textwidth]{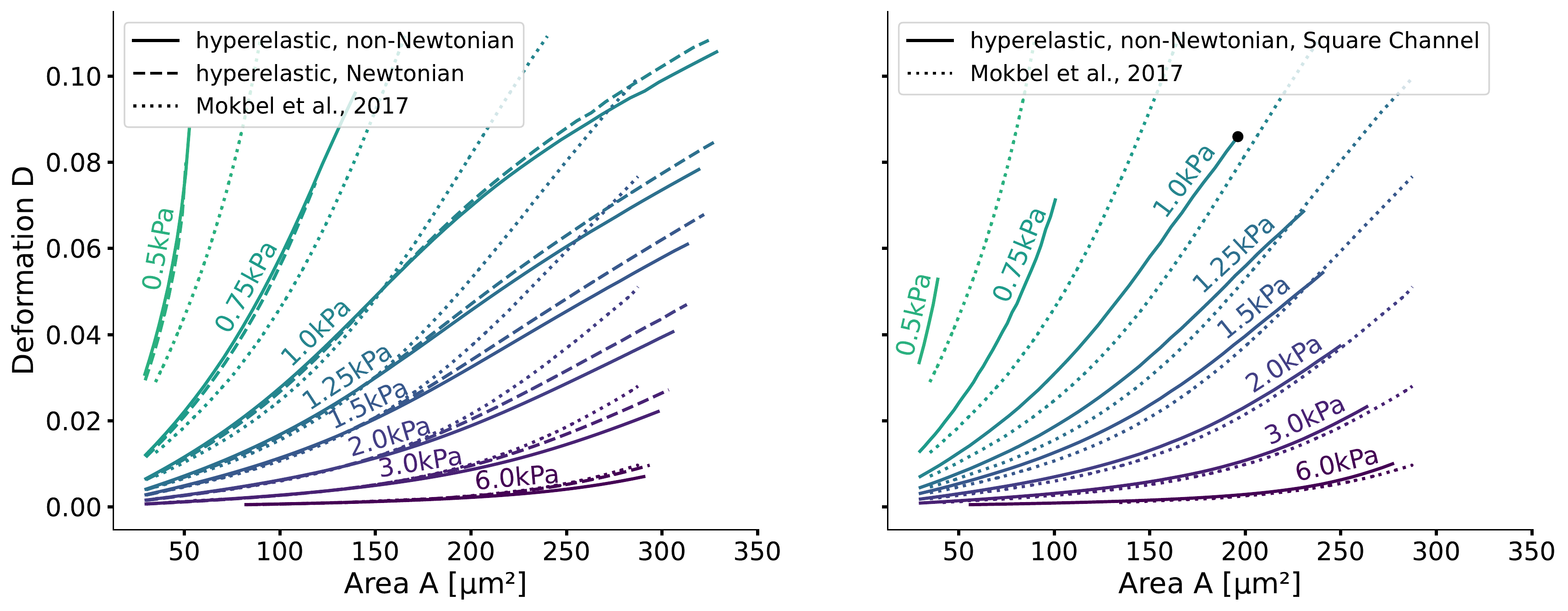} 
    \caption{
        {\bf Comparison of the stationary deformation for the linear and hyperelastic cells:}
    	Isoelasticity lines for different Young's moduli $E \in [0.5, 0.75, 1.0, 1.25, 1.5, 2.0, 3.0, 6.0]\SI{}{kPa}$ for the linear elastic material model without shear-thinning (dotted, see \cite{Mokbel2017-pb}), the new hyperelastic material model with shear-thinning (solid) and without shear-thinning (dashed). 
    	On the {\bf (left)} for the cylindrical (2D-axisymmetric) channel.
    	On the {\bf (right)} for the square (full 3D) channel geometry.  
    	The black dot corresponds to the simulation shown in \cref{fig:overview}.
    	Parameters:
		$L = \SI{20}{\um}$,
		$Q = \SI{0.04}{\ul.s^{-1}}$,
		buffer characteristics of \ccb{} for the shear-thinning isoelasticity lines, $\eta = \SI{6}{mPa.s}$ for the apparent viscosity of \ccb{} in the Newtonian case. 
		Contours of \cite{Mokbel2017-pb} extracted from \cite{Wittwer2020-eb}.
    }
    \label{fig:materialmodels}
\end{figure}

Next, we validate that the resulting lookup tables are -- like the linear elastic LUT -- bijective (no isoelasticity lines cross each other) and the scaling holds for the hyperelastic material model, the non-Newtonian shear-thinning buffer as well as the square channel geometry. 
Isoelasticity lines for different Young's moduli in the cylindrical channel with and without a shear-thinning buffer are shown \cref{fig:materialmodels} (left). 
In \cref{fig:materialmodels} (right), the same is shown for the square channel with a shear-thinning buffer only. 
In both plots, the isoelasticity lines from \citep{Mokbel2017-pb} for the linear material model without shear thinning and cylindrical channel are added for comparison (dotted lines). 
All three lookup tables are bijective, meaning that for each area-deformation pair, there is a unique apparent Young's modulus associated. 
For small deformations (read small strains), the linear elastic LUT does agree with the hyperelastic material based LUTs. 
For stiff beads, this is the case for almost all cell sizes.
Reducing the cell stiffness leads to increased divergences of the isoelasticity lines for larger beads. 
For soft beads, the isoelasticity lines diverge already for small bead size. 
This holds true for the cylindrical channel simulations as well as for the square channel simulations. 
In the case of a cylindrical channel, the isoelasticity lines flatten out for Young's moduli around $\SI{1}{kPa}$ for larger cells, whereas the linear elastic material predicts a higher increase in deformation.
The square channel lookup table (\cref{fig:materialmodels} (right)) shows a similar pattern. 
There, the isoelasticity lines agree for stiff cells. 
For softer cells, the isoelasticity lines diverge already at smaller cell sizes and the deformations are larger.
For larger cells with a Young's modulus $\leq \SI{1.5}{kPa}$, the isoelasticity lines seem to flatten out too, but the range in terms of numerically stable simulations is smaller, resulting in a smaller lookup table than in the cylindrical channel. 
Additionally, for the cylindrical channel, we investigate the influence of shear-thinning on the isoelasticity lines by comparing to the deformations in a Newtonian fluid with a fixed apparent viscosity $\eta = \SI{6}{mPa.s}$ (see \citep{Herold2017-ir} for derivation from a shear-thinning fluid).
The resulting isoelasticity lines are shown again in \cref{fig:materialmodels} (left)) as dashed lines. 
For smaller cells, shear thinning is negligible for all cell elasticities.
Again, for the larger cells, a Newtonian fluid results in a higher deformation. 
This leads to an underestimation of the apparent Young's modulus for larger cells.
Overall, the agreement between the linear elastic and hyperelastic isoelasticity lines for small strains validates the new numerical results.

\begin{figure}[tp]
    \centering
    \includegraphics[width=.75\textwidth]{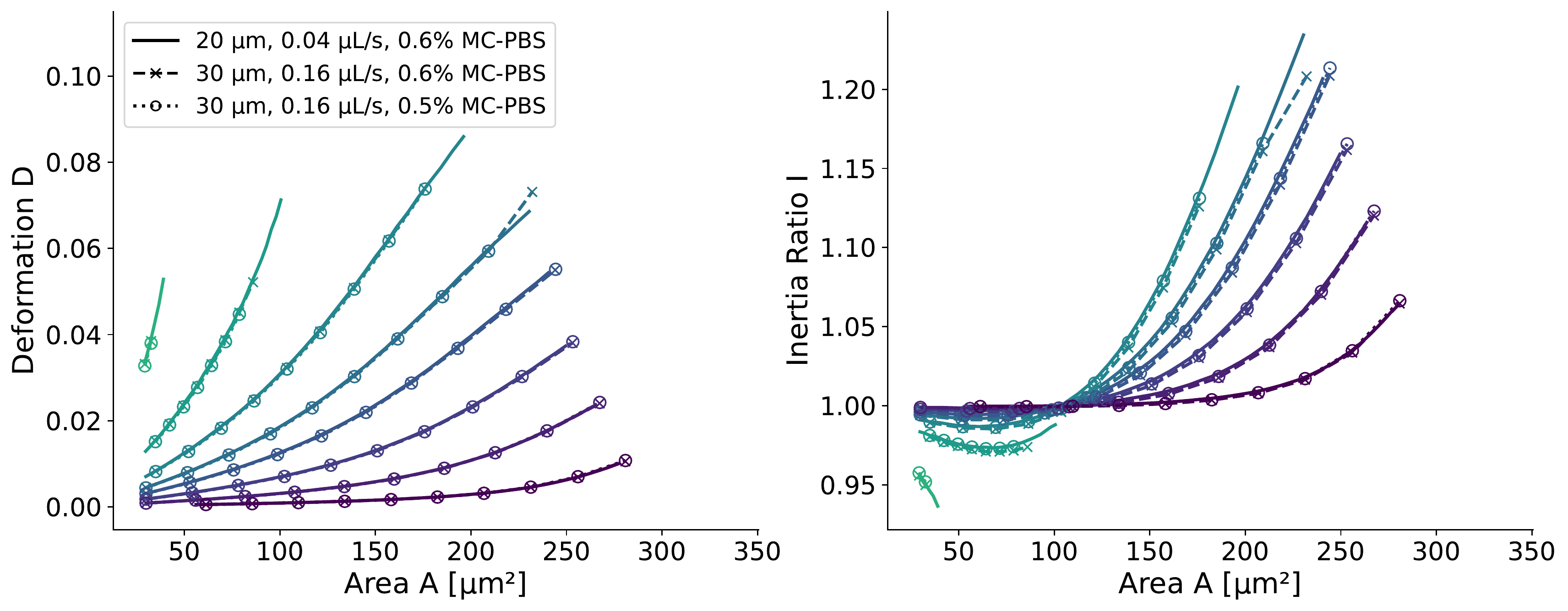}
	\caption{
		{\bf Scaling of the square channel lookup table:}
		Isoelasticity lines of the new lookup table (solid) in the square channel compared to rescaled data with a wider channel geometry and a higher flow rate (dashed) and additionally a different fluid (dotted).
		The isoelasticity lines and colouring is the same as in \cref{fig:2dto3d} (left).
		The Young's modulus $E^\prime$ and the projected cell area $A^\prime$ were scaled to match the new channel geometry, flow rate and apparent viscosities.
		On the {\bf (left)} for the deformation, on the {\bf (right)} for the inertia ratio. 
	}
	\label{fig:scaling}
\end{figure}

Validating the scaling discussed above \cite{Mietke2015-cj} is non-trivial, as we include the non-linear shear-thinning behaviour of \ccb{} and hyperelastic neo-Hookean material model for the cell and consider the full three-dimensional geometry.
The new lookup tables are based on a channel side length $L = \SI{20}{\um}$, a flow rate of $Q = \SI{0.04}{\ul.s^{-1}}$ and an apparent viscosity of \ccb{} which results in $\eta = \SI{6}{mPa.s}$.
Here, to validate the scaling numerically, we perform two \emph{in silico} experiments in square channels with a side length of $L^\prime = \SI{30}{\um}$, a flow rate of $Q^\prime = \SI{0.16}{\ul.s^{-1}}$ and with \cc{} and \ccb{}, resulting in apparent viscosities $\eta^\prime_1 = \SI{5.64}{mPa.s}$ and $\eta^\prime_2 = \SI{4.76}{mPa.s}$.
We scaled the Young's moduli and derived the deformation values from the adapted simulations.
After rescaling the area $A^\prime$, the isoelasticity lines in \cref{fig:scaling} (left) agree perfectly except at the boundary, indicating that the scaling still holds at least for the most part of the LUT and only at the boundary it might become inaccurate. 

The ratio between the two axial second  moments of area is another dimensionless integral quantity to describe cell deformation, introduced in \citep{Mokbel2017-pb}.
We use the definition of this \emph{inertia ratio} $I$ implemented in dclab \citep{dclab} which for horizontally symmetric shapes is defined by 
\begin{align}
    I &= \sqrt{\frac{I_{yy}}{I_{xx}}}, \\
    I_{xx} &= \int_A (y-y_b)^2 dA, \\
    I_{yy} &= \int_A (x-x_b)^2 dA,
\end{align}
where $A$ is the projected cell area and $(x_b,y_b)$ is the cell barycentre \citep{Herbig2018-pq}.
Similar to isoelasticity lines in the area-deformation plot, one can plot the isoelasticity lines in the area-inertia ratio space shown in \cref{fig:2dto3d} (right). 
Cells which are more stretched in the direction of the fluid flow (prolate) have an inertia ratio $I >1$.
An inertia ratio $I < 1$ means that the cells are more compressed in the flow direction and elongated perpendicular, thus resemble an oblate. 
The area-inertia ratio lookup table is not bijective, as already reported in \cite{Mokbel2017-pb}.
Around an area of \SI{108}{\um^2} the isoelasticity lines cross each other, meaning that in this region a unique lookup for the apparent Young's modulus is not possible. 
But, the scaling does hold too for the inertia ratio as seen in \cref{fig:2dto3d} (right).
Thus, the presented lookup tables can be used for deriving uniquely the apparent Young's modulus for a broad range of experimental setups from the area-deformation space and to some extent from the area-inertia ratio space. 

\subsection*{The New Lookup Tables and the Geometric Influence}
\begin{figure}[tp]
    \centering 
    \includegraphics[width=.75\textwidth]{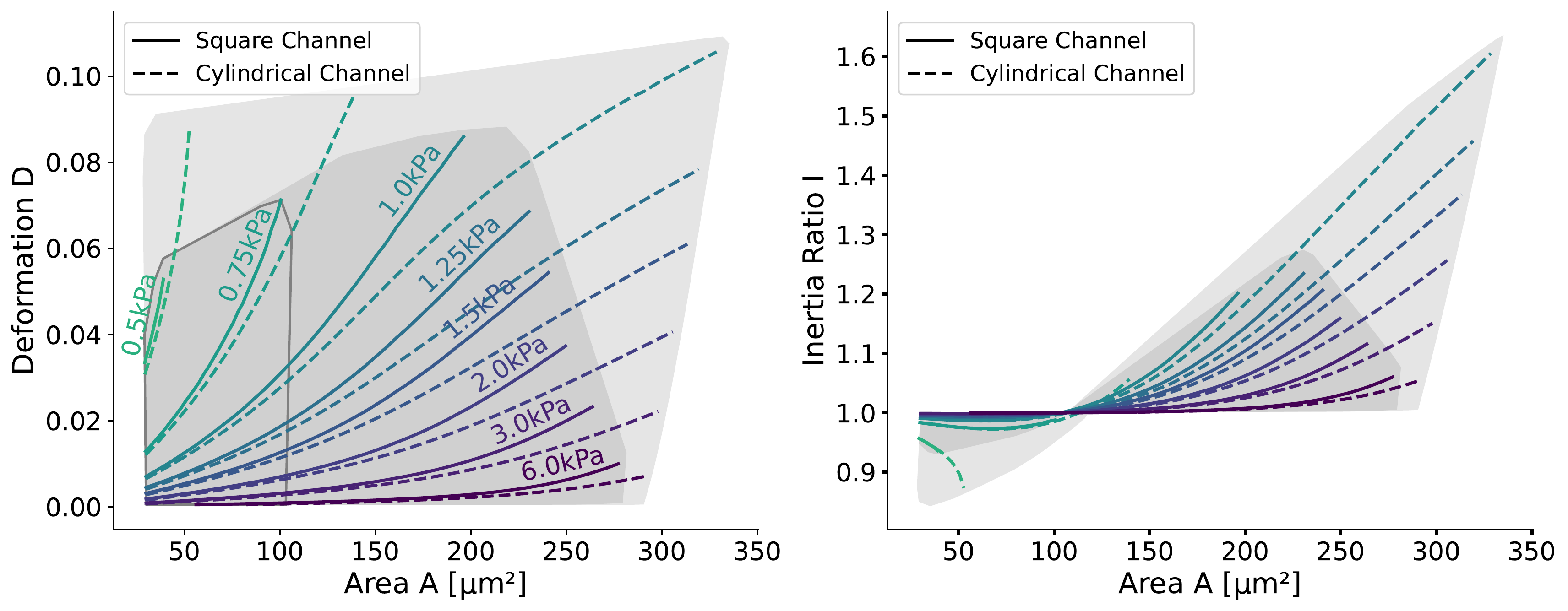} 
    \caption{
        {\bf Comparison of the new lookup tables in the cylindrical and square channel:} 
    	{\bf (left)} 
    	Isoelasticity lines of cells in the square channel (solid) and cylindrical channel (dotted). 
    	The region of all the simulations are indicated by the shaded areas. 
    	{\bf (right)} 
    	Inertia ratio of the two lookup tables: solid lines for the square channel and dotted lines for the cylindrical channel. 
    	The area with an inertia ratio $I < 1$ is indicated by the enclosed area with the gray line on the left.
    	}
    \label{fig:2dto3d}
\end{figure}

Having validated the numerical simulations and the scaling, we can construct the final lookup tables based on many simulations.
As already described above, the lookup tables are based on simulations with a channel side length $L = \SI{20}{\um}$, a flow rate of $Q = \SI{0.04}{\ul.s^{-1}}$ and \ccb{}.
The cylindrical channel simulations are numerically more stable as the confinement -- the ratio between the cell radius and channel width -- is smaller, i.e., the channel width is larger to account for the changed pressure drop.
For more details of the concept of \emph{equivalent channel} between the square and cylindrical channel, see \cite{Mietke2015-cj}.
\Cref{fig:2dto3d} shows the region of all simulations for the two new lookup tables in the area-deformation space as well as area-inertia ratio space.
The space of the cylindrical lookup table spans $[\SI{28}{\um^2}, \SI{335}{\um^2}] \times [\SI{0.377}{kPa}, \SI{23.753}{kPa}]$ and the square channel $[\SI{29}{\um^2}, \SI{281}{\um^2}] \times [\SI{0.469}{kPa}, \SI{27.669}{kPa}]$.
Since the area is not conserved and not all radius-Young's modulus pairs are numerically stable, the resulting LUTs are not rectangular.
Again, we show the isoelasticity lines for the same Young's moduli as in \cref{fig:materialmodels}. 
The area where the inertia ratio of the square channel is $I < 1$ is enclosed in \cref{fig:2dto3d} (left) by the grey line. 

The channel geometry does have observable influence on the deformations. 
On the technical side, the simulations are more stable in the axisymmetric setting as discussed above, resulting in a larger lookup table.
For small cells as well as stiff cells, the isoelasticity lines from the two new LUTs overlay. 
But for larger and softer cells, the isoelasticity lines of the cylindrical and square channel diverge.
As the deformation is smaller in the cylindrical channel, the cylindrical lookup table underestimates the apparent Young's modulus.

\subsection*{Applying the new LUTs to Measurements}
\begin{figure}[tp]
    \centering
    \includegraphics[width=.75\textwidth]{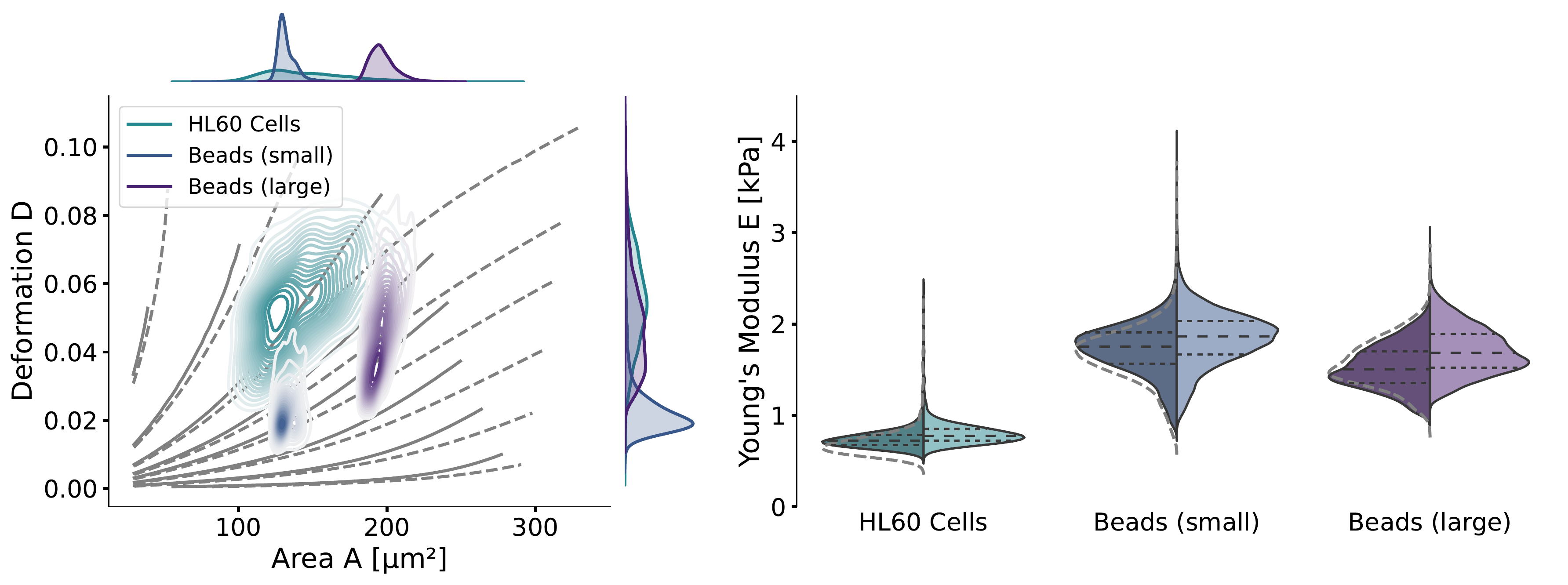} 
    \caption{
    	{\bf RT-DC measurements of PAAm beads and HL60:} 
    	{\bf (left)}
    	The distributions of PAAm beads with two different sizes and HL60 cells from separate measurements (by bivariate kernel density estimation).
    	The marginal distributions for area and deformation are shown on the top and on the right.
    	The isoelasticity lines from \cref{fig:2dto3d} (left) are shown in grey. 
    	{\bf (right)}
    	Violin plots of the estimated Young's modulus from the three measurements. 
    	The resulting Young's modulus distribution with the cylindrical LUT is on the left in each violin, for the square channel LUT on the right. 
    	The dashed lines within the violins are the quartiles (25\% / 50\% /  75\%).
    	The grey dashed lines indicate the Young's modulus distribution from the cylindrical LUT of \cite{Mokbel2017-pb}.
    }
    \label{fig:beads}
\end{figure}

With the validated new lookup tables, it is possible to extract the apparent Young's modulus from RT-DC measurements. 
To demonstrate this, we measured poly-acrylamide (PAAm) hydrogel beads \citep{Girardo2018-yk} and HL60 cells (see Methods). 
The kernel density estimations of the measured bead and cell distributions are shown in \cref{fig:beads} (left).
All three measurements fall within the region of definition of both new LUTs, with a few sample points outside the square channel LUT (resulting in omitted data points). 
The resulting apparent Young's modulus is shown in \cref{fig:beads} (right). 
The cylindrical LUT predicts similar elasticity values and distribution than the linear elastic LUT of \cite{Mokbel2017-pb}. 
For small beads, the new cylindrical channel LUT predicts a higher Young's modulus compared to the linear elastic LUT, whereas for the larger beads the Young's moduli decrease (see the gray dashed lines in \cref{fig:beads} (right)). 
This is due to the non-linear material response discussed above and seen in the flattening isoelasticity lines. 
For the HL60 cells which have a higher deformation than the small beads, the stiffness increases too compared to the linear elastic LUT. 
Comparing the two new LUTs, the square channel LUT predicts higher Young's moduli for all three measurements.
This holds true compared to the old linear elastic material model in the cylindrical channel. 
At the same time, the shape of the distribution changes compared to the cylindrical LUT, supporting the importance of the channel geometry. 

\subsection*{Strain and stresses of the cells}
\begin{figure}[tp]
    \centering
    \includegraphics[width=.75\textwidth]{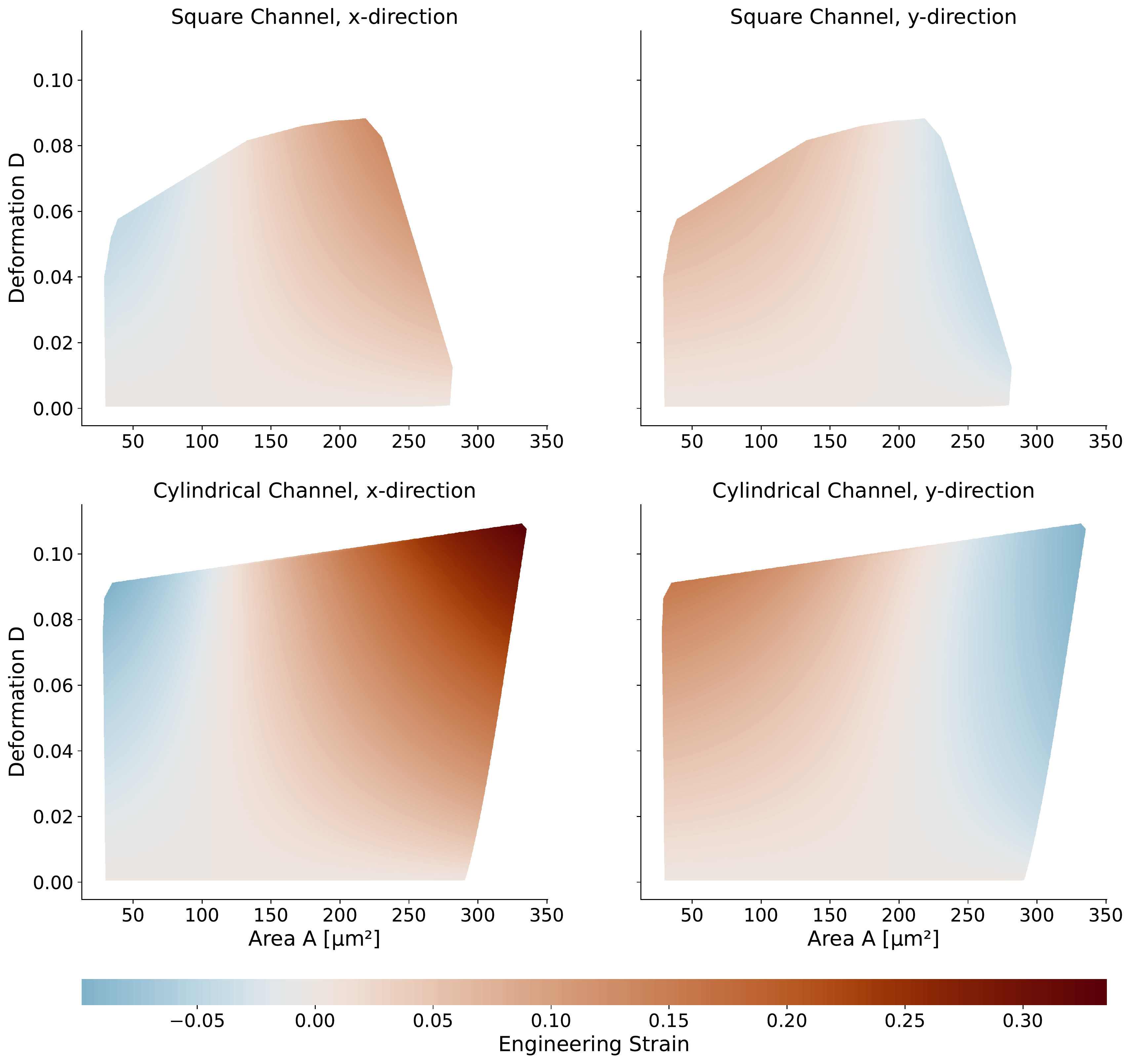} 
    \caption{
        {\bf Engineering strain of the cell shapes:}
        {\bf (top)}
        Strain of the cells in the square channel. On the left in the x-direction (flow direction) and on the right in y-direction (perpendicular to the flow direction)
        {\bf (bottom)}
        Same strain measure but for the cells in the cylindrical channel.
        }
    \label{fig:strains}
\end{figure}

Next, we will look at the strain and stresses on the cell surface.
Deformation as a measure of strain approximates the resulting shapes by a single number.
In \cref{fig:strains}, we plot the strains along the horizontal x-axis and the vertical y-axis to get a better understanding of the deformed shapes. 
Here we use the engineering strain defined as $(l-l_0) / l_0$ and $(h-h_0) / h_0$ where $l$ and $h$ are the maximum extensions of the cell in x and y direction.
$l_0$ and $h_0$ are the diameter of the cell in the undeformed state with $l_0 = h_0 = 2 \cdot R$. 
Small cells are horizontally compressed but elongated vertically. 
Larger cells are exposed to higher shear force, resulting in a more stretched shape along the flow direction and compressed in the y-axis. 
The strains of the cells in the square channel are in the range of [-4.5\%, 13\%], in the cylindrical channel in the range of [-9.9\%, 33.6\%]. 

\begin{figure}[tp]
    \centering
    \includegraphics[width=.75\textwidth]{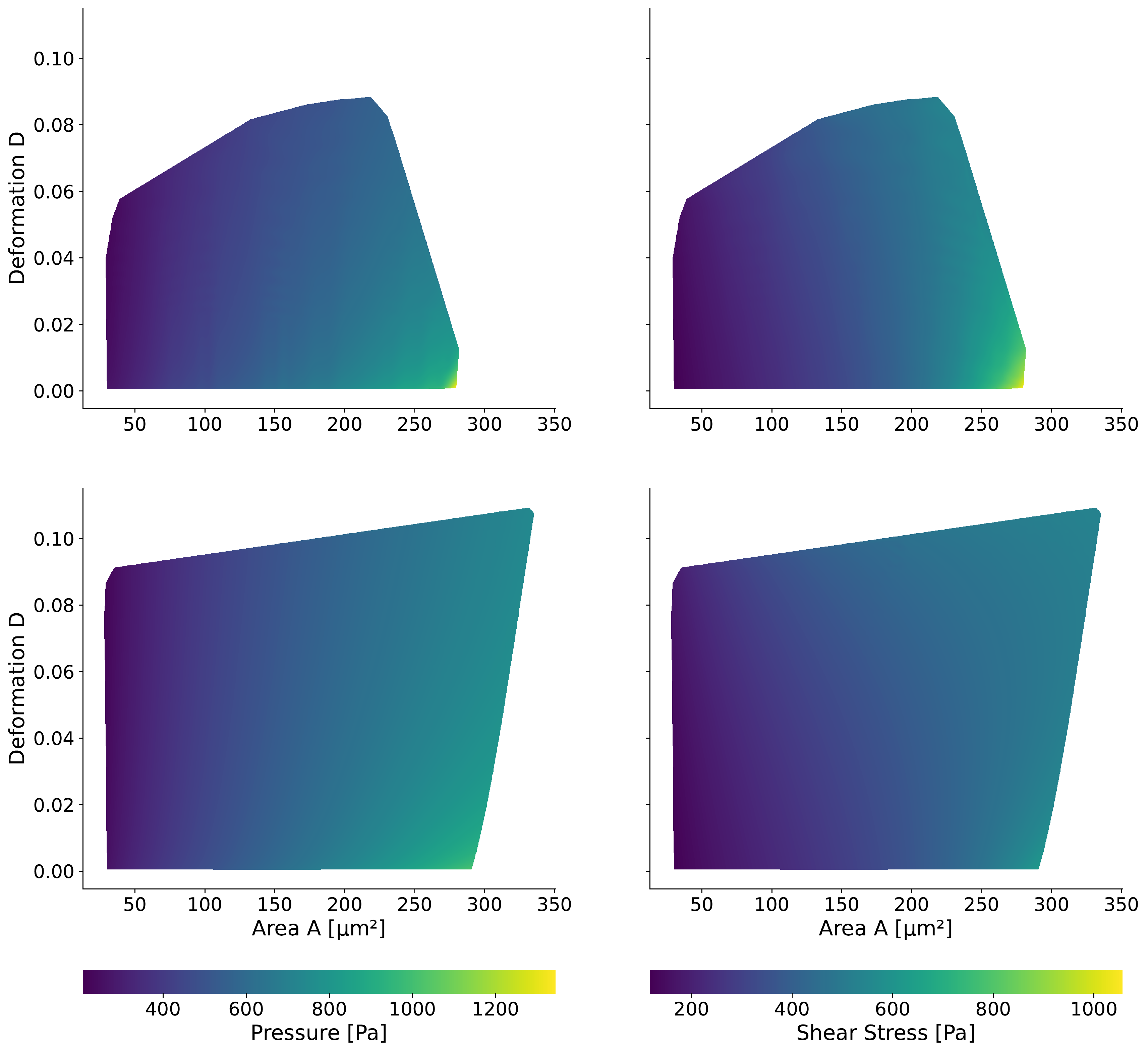} 
    \caption{
        {\bf Maximum surface stresses on the cells:}
        The maximum pressure (left) and shear stress (right) on the cell contour for the square channel LUT (top) and cylindrical channel (bottom). 
        The pressure is normalized by the average surface pressure $p_\text{avg}$.
    }
    \label{fig:stresses}
\end{figure}

The maximum stress components acting on the cells are shown in \cref{fig:stresses} for the square and cylindrical channel.
The shear stresses are in the range of [\SI{117}{Pa}, \SI{1058}{Pa}] in the square channel and [\SI{118}{Pa}, \SI{624}{Pa}] in the cylindrical channel, again increasing with the cell size. 
The largest cells with a large Young's modulus are exposed to the highest shear stress, as their surface is closest to the channel wall. 
The maximum pressure is in the range of [\SI{226}{Pa}, \SI{1346}{Pa}] in the square channel and [\SI{220}{Pa}, \SI{1022}{Pa}] in the cylindrical channel, respectively. 
We subtract the average surface pressure $p_\text{avg}$ again (see above).

\subsection*{Volume estimation error}
\begin{figure}[tp]
    \centering
    \includegraphics[width=.5\textwidth]{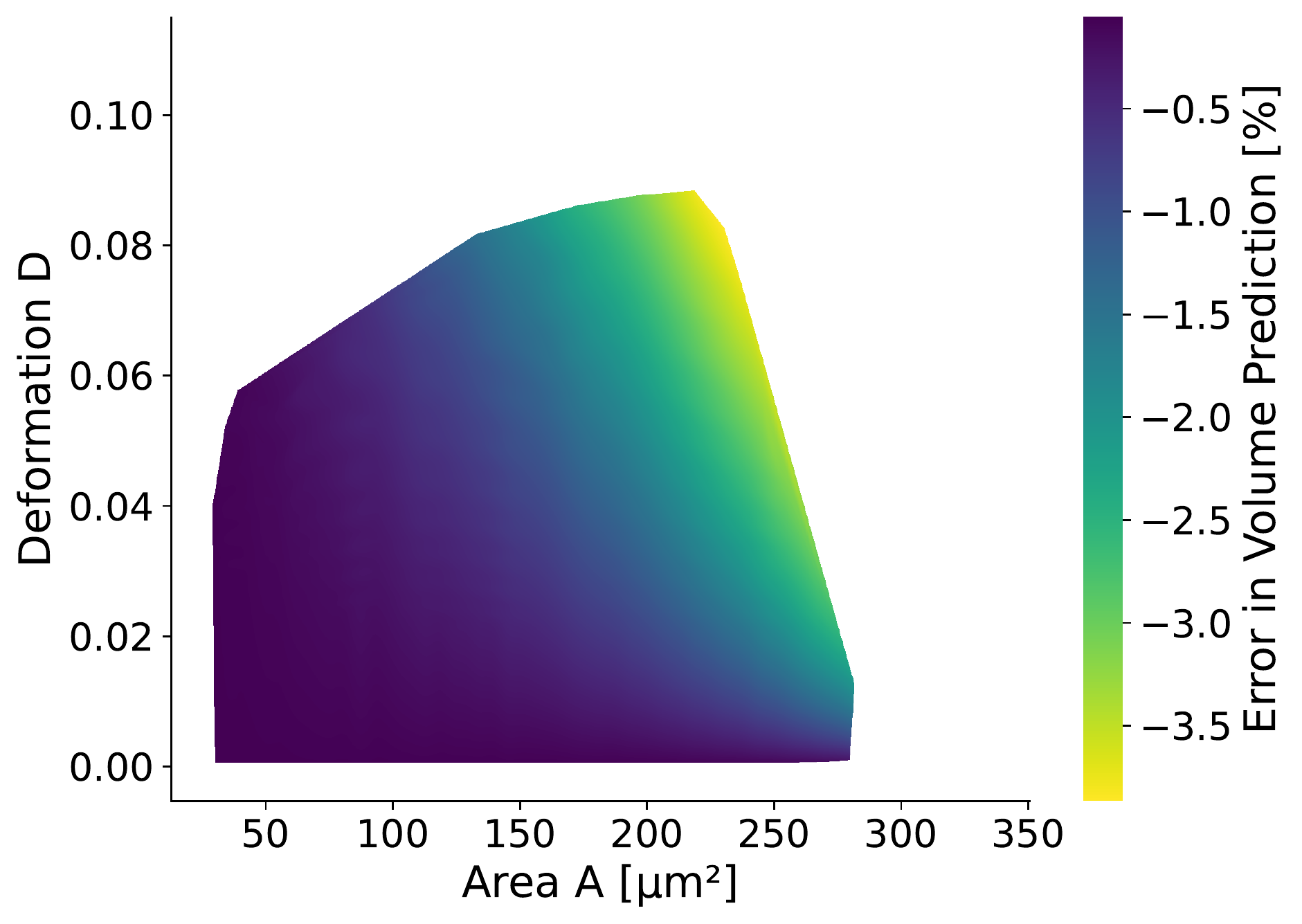} 
	\caption{
	    {\bf Relative error of volume prediction:}
	    The volume approximation from the cell contours assumes rotational symmetry to calculate the volume. 
	    Due to the square channel, the volume is overestimated up to 3.5\% depending on the projected area and deformation for the stationary deformation.}
	\label{fig:volerror}
\end{figure}

The volume of cells flowing through the microfluidic chip is a valuable cell feature. 
It is approximated by revolving the detected contour of the stationary shape and pixelation corrected similar to the cell deformation and area (see Methods).
The rotational-symmetry assumption is only valid for small cells, perceiving the almost axisymmetric flow field in the centre of the channel. 
\Cref{fig:volerror} shows the relative error based on measured area and deformation. 
Interestingly, the error is only in the range of $[-3.87\%, -0.06\%]$. 
The volume of small cells is underestimated only slightly compared to the actual cell volume independently of the deformation and thus apparent Young's modulus. 
For larger cells, the influence of the deformation on the volume error increases.

\section*{Discussion \& Conclusion}


In this work, we present two new lookup tables for RT-DC for cylindrical and square channels based on the neo-Hookean hyperelastic material model.
We validate the numerical model by comparing to the linear elastic LUT of \cite{Mokbel2017-pb} in the small strain regime. 
We discuss the influence of different material models as well as the shear-thinning behaviour of the fluid as well as the channel geometry and reveal different material responses.
We find that the scaling still holds for the hyperelastic material even with a shear-thinning non-Newtonian fluid.
In consequence, the proposed LUTs can be used for a vast array of different channel sizes, buffer viscosities and flow rates. 
Shear-thinning does affect the isoelasticity lines for large cells only.
Compared to the existing linear elastic material LUT, the more accurate hyperelastic square channel LUT predicts stiffer cells, especially for small highly deformed cells. 
However, the square channel LUT has a limited range of possible cell sizes and deformations due to the smaller region of numerical stable simulations. 
On the other hand, the cylindrical channel LUT has a similar size as the linear elastic material LUT.
Depending on the position of the measurements in the area-deformation space, the distribution of the Young's modulus is altered for both proposed LUTs compared to the existing one.
For both proposed LUTs, we report the strain and surface stresses of the cells depending on the projected cell area and deformation. 
Both new lookup tables are available in dclab \cite{dclab} 
and freely available in \cite{Wittwer2022-pw}, 
extending the predictive power to derive the apparent Young's modulus of biological cells.
At the same time, the proposed LUTs complement the work in \cite{Wittwer2022-mj} to derive the viscoelastic response of biological cells in a square RT-DC.
A lookup based on area and deformation ignores the actual shape of the measured cells.
Extending the search space by considering the contour shapes would considerably improve the predictive power of RT-DC and could take into account different material models for different cell types and beads,
However, this leads to a significant increase in parameters and is thus left for future studies.

\section*{Acknowledgements}
Simulations were performed at the Center for Information Services and High Performance Computing (ZIH) at TU Dresden. L.W. acknowledges the COMSOL support staff for their support. 
S.A. acknowledges support from the German Research Foundation DFG (grant AL1705/3-2) and 
support from the Saxon Ministry for Science and Art (SMWK MatEnUm-2 and EUProfil).

\appendix

\bibliographystyle{elsarticle-num}
\bibliography{hyperelastic_lut_rtdc}

\begin{thebibliography}{10}
\expandafter\ifx\csname url\endcsname\relax
  \def\url#1{\texttt{#1}}\fi
\expandafter\ifx\csname urlprefix\endcsname\relax\def\urlprefix{URL }\fi
\expandafter\ifx\csname href\endcsname\relax
  \def\href#1#2{#2} \def\path#1{#1}\fi

\bibitem{Toepfner2018-al}
N.~Toepfner, C.~Herold, O.~Otto, P.~Rosendahl, A.~Jacobi, M.~Kr{\"a}ter,
  J.~St{\"a}chele, L.~Menschner, M.~Herbig, L.~Ciuffreda,
  L.~Ranford-Cartwright, M.~Grzybek, {\"U}.~Coskun, E.~Reithuber, G.~Garriss,
  P.~Mellroth, B.~Henriques-Normark, N.~Tregay, M.~Suttorp, M.~Bornh{\"a}user,
  E.~R. Chilvers, R.~Berner, J.~Guck, {Detection of human disease conditions by
  single-cell morpho-rheological phenotyping of blood}, Elife 7 (2018) e29213.

\bibitem{Koch2017-ji}
M.~Koch, K.~E. Wright, O.~Otto, M.~Herbig, N.~D. Salinas, N.~H. Tolia, T.~J.
  Satchwell, J.~Guck, N.~J. Brooks, J.~Baum, {Plasmodium falciparum
  erythrocyte-binding antigen 175 triggers a biophysical change in the red
  blood cell that facilitates invasion}, Proc. Natl. Acad. Sci. U. S. A.
  114~(16) (2017) 4225--4230.
\newblock \href {https://doi.org/10.1073/pnas.1620843114}
  {\path{doi:10.1073/pnas.1620843114}}.

\bibitem{Kubankova2021-uu}
M.~Kub{\'a}nkov{\'a}, B.~Hohberger, J.~Hoffmanns, J.~F{\"u}rst, M.~Herrmann,
  J.~Guck, M.~Kr{\"a}ter, Physical phenotype of blood cells is altered in
  {COVID-19}, Biophys. J. 120~(14) (2021) 2838--2847.
\newblock \href {https://doi.org/10.1016/j.bpj.2021.05.025}
  {\path{doi:10.1016/j.bpj.2021.05.025}}.

\bibitem{Otto2015-hf}
O.~Otto, P.~Rosendahl, A.~Mietke, S.~Golfier, C.~Herold, D.~Klaue, S.~Girardo,
  S.~Pagliara, A.~Ekpenyong, A.~Jacobi, M.~Wobus, N.~T{\"o}pfner, U.~F. Keyser,
  J.~Mansfeld, E.~Fischer-Friedrich, J.~Guck, {Real-time deformability
  cytometry: on-the-fly cell mechanical phenotyping}, Nat. Methods 12~(3)
  (2015) 199--202, 4 p following 202.

\bibitem{Mietke2015-cj}
A.~Mietke, O.~Otto, S.~Girardo, P.~Rosendahl, A.~Taubenberger, S.~Golfier,
  E.~Ulbricht, S.~Aland, J.~Guck, E.~Fischer-Friedrich, Extracting cell
  stiffness from {Real-Time} deformability cytometry: Theory and experiment,
  Biophys. J. 109~(10) (2015) 2023--2036.

\bibitem{Mokbel2017-pb}
M.~Mokbel, D.~Mokbel, A.~Mietke, N.~Tr{\"a}ber, S.~Girardo, O.~Otto, J.~Guck,
  S.~Aland, {Numerical Simulation of {Real-Time} Deformability Cytometry To
  Extract Cell Mechanical Properties}, ACS Biomater Sci Eng 3~(11) (2017)
  2962--2973.

\bibitem{dclab}
P.~M{\"u}ller, M.~Herbig, E.~O'Connell, P.~Rosendahl, M.~Schl{\"o}gel, J.~Guck,
  \href{https://github.com/ZELLMECHANIK-DRESDEN/dclab}{dclab version 0.34.1:
  Python library for the post-measurement analysis of real-time deformability
  cytometry data sets [software]} (2015).
\newline\urlprefix\url{https://github.com/ZELLMECHANIK-DRESDEN/dclab}

\bibitem{Herold2017-ir}
C.~Herold, \href{http://arxiv.org/abs/1704.00572}{{Mapping of Deformation to
  Apparent Young's Modulus in {Real-Time} Deformability Cytometry}} (Apr.
  2017).
\newblock \href {http://arxiv.org/abs/1704.00572} {\path{arXiv:1704.00572}}.
\newline\urlprefix\url{http://arxiv.org/abs/1704.00572}

\bibitem{Fregin2019-up}
B.~Fregin, F.~Czerwinski, D.~Biedenweg, S.~Girardo, S.~Gross, K.~Aurich,
  O.~Otto, {High-throughput single-cell rheology in complex samples by dynamic
  real-time deformability cytometry}, Nat. Commun. 10~(1) (2019) 415.

\bibitem{Schuster2021-hx}
R.~Schuster, O.~Marti, {Simulation of cell deformation inside a microfluidic
  channel to identify parameters for mechanical characterization of cells}, J.
  Phys. D Appl. Phys. 54~(12) (2021) 125401.

\bibitem{Wittwer2022-mj}
L.~D. Wittwer, F.~Reichel, S.~Aland, Numerical simulation of deformability
  cytometry - transport of a biological cell through a microfluidic channel,
  in: S.~Becker, A.~Kuznetsov, F.~de~Monte, G.~Pontrelli, D.~Zhao (Eds.),
  Modeling of Mass Transport Processes in Biological Media, Elsevier, 2022.

\bibitem{Girardo2018-yk}
S.~Girardo, N.~Tr{\"a}ber, K.~Wagner, G.~Cojoc, C.~Herold, R.~Goswami,
  R.~Schl{\"u}{\ss}ler, S.~Abuhattum, A.~Taubenberger, F.~Reichel, D.~Mokbel,
  M.~Herbig, M.~Sch{\"u}rmann, P.~M{\"u}ller, T.~Heida, A.~Jacobi, E.~Ulbricht,
  J.~Thiele, C.~Werner, J.~Guck, Standardized microgel beads as elastic cell
  mechanical probes, J. Mater. Chem. B Mater. Biol. Med. 6~(39) (2018)
  6245--6261.

\bibitem{Urbanska2018-hs}
M.~Urbanska, P.~Rosendahl, M.~Kr{\"a}ter, J.~Guck,
  \href{http://dx.doi.org/10.1016/bs.mcb.2018.06.009}{High-throughput
  single-cell mechanical phenotyping with real-time deformability cytometry},
  Methods Cell Biol. 147 (2018) 175--198.
\newblock \href {https://doi.org/10.1016/bs.mcb.2018.06.009}
  {\path{doi:10.1016/bs.mcb.2018.06.009}}.
\newline\urlprefix\url{http://dx.doi.org/10.1016/bs.mcb.2018.06.009}

\bibitem{Herbig2018-gl}
M.~Herbig, M.~Kr{\"a}ter, K.~Plak, P.~M{\"u}ller, J.~Guck, O.~Otto,
  \href{http://dx.doi.org/10.1007/978-1-4939-7346-0_15}{{Real-Time}
  deformability cytometry: {Label-Free} functional characterization of cells},
  Methods Mol. Biol. 1678 (2018) 347--369.
\newblock \href {https://doi.org/10.1007/978-1-4939-7346-0\_15}
  {\path{doi:10.1007/978-1-4939-7346-0\_15}}.
\newline\urlprefix\url{http://dx.doi.org/10.1007/978-1-4939-7346-0_15}

\bibitem{Wittwer2020-eb}
L.~D. Wittwer, P.~M{\"u}ller, D.~Mokbel, M.~Mokbel, J.~Guck, S.~Aland, Finite
  element simulation data for the computation of the young's modulus in
  real-time deformability cytometry (Apr. 2020).
\newblock \href {https://doi.org/10.6084/M9.FIGSHARE.12155064.V2}
  {\path{doi:10.6084/M9.FIGSHARE.12155064.V2}}.

\bibitem{Herbig2018-pq}
M.~Herbig, A.~Mietke, P.~M{\"u}ller, O.~Otto, {Statistics for real-time
  deformability cytometry: Clustering, dimensionality reduction, and
  significance testing}, Biomicrofluidics 12~(4) (2018) 042214.

\bibitem{Wittwer2022-pw}
L.~D. Wittwer, P.~M{\"u}ller, F.~Reichel, S.~Aland, J.~Guck, {Hyperelastic
  Lookup Table for Real-Time Deformability Cytometry ({RT-DC})} (Aug. 2022).
\newblock \href {https://doi.org/10.6084/m9.figshare.20630940}
  {\path{doi:10.6084/m9.figshare.20630940}}.

\end{thebibliography}

\end{document}